\documentclass[reprint, secnumarabic,amssymb, nobibnotes, aps, prl]{revtex4-1}

\usepackage{color}
\usepackage{epsfig, graphics, graphicx, subfigure, wrapfig, lipsum}
\usepackage{verbatim}
\usepackage{dcolumn}
\usepackage{bm}
\usepackage{amsmath, braket}
\usepackage{array}
\usepackage{xr}

\newcolumntype{X}[1]{>{\centering\arraybackslash\hspace{0pt}}p{#1}}
\newcolumntype{M}[1]{ >{\centering\arraybackslash}m{#1}}
\newcommand{\roml}[1]{\lowercase\expandafter{\romannumeral #1\relax}}
\newcommand{\romu}[1]{\uppercase\expandafter{\romannumeral #1\relax}}

\begin{document}

\title{Non-ballistic Thermal Transport in Carbon Nanotubes}

\author{Ankit Jain}
\email{a\_jain@iitb.ac.in}
\affiliation{Mechanical Engineering Department, IIT Bombay, India}
\date{\today}

\begin{abstract}
{The thermal transport properties of single-wall carbon nanotubes (SWCNTs) are re-investigated using the iterative solution of the Boltzmann transport equation by including four-phonon scattering. Using only three-phonon scattering, the flexural and twisted phonon modes are found to remain non-scattered via Umklapp processes, resulting in the literature-reported divergence of thermal conductivity with tube length. However, with four-phonon scattering, while longitudinal modes remain unaffected, the otherwise non-scattered transverse modes undergo Umklapp scattering and result in a non-ballistic transport, thus settling a decades-long debate on the length dependence of thermal conductivity of SWCNTs. The predicted thermal conductivity of SWCNTs using both three- and four-phonon scatterings is 3700 W/m-K at 300 K and stays highest amongst all known materials.}

\end{abstract}
\maketitle

The understanding of thermal transport physics of carbon nanotubes has remained an open research problem to date \cite{hone1999, kim2001,pop2006,shiomi2006,donadio2007,chang2008,lindsay2009,lee2017,bruns2020,barbalinardo2021}. Experimentally, on the one hand, the thermal transport in single-wall carbon nanotubes (SWCNTs) is measured to be non-diffusive with divergence of thermal conductivity ($\kappa$) for tube lengths of up to 1 mm \cite{chang2008, lee2017} as suggested by the  Fermi, Pasta, Ulam (FPU), and Tsingou model \cite{fermi1955}, on the other hand, the $\kappa$ is recently reported to converge for tube lengths of only 10 $\mu$m \cite{liu2017}; thus, highlighting challenges in the experimental measurements and interpretation of thermal transport results for SWCNTs \cite{li2017}.

Earlier theoretical studies, based on phonon scattering selection rules,  suggested non-scattering of long wavelength flexural acoustic and twisted/rotational acoustic phonon modes (collectively referred to as transverse modes, hereafter) via the Umklapp three-phonon processes \cite{lindsay2009, lindsay2010a}. This is indeed confirmed numerically from mode-dependent phonon properties obtained by employing the iterative solution of the Boltzmann transport equation (BTE) where $\kappa$ is found to diverge in the absence of phonon-boundary scattering \cite{lindsay2009}. These theoretical predictions and numerical mode-dependent phonon properties are, however, obtained by considering only three-phonon phonon processes, and it is not clear if long wavelength transverse phonon remains unscattered when higher-order four-phonon processes are accounted for \cite{lindsay2009, bruns2020}. Other computational approaches based on molecular dynamics simulations can naturally include phonon anharmonicity to the highest order. However, these simulations remained inconclusive for SWCNTs with equilibrium molecular dynamics suggesting converged $\kappa$ \cite{donadio2007, thomas2010} and the direct molecular dynamics suggesting length dependence of $\kappa$ for up to at least 10 $\mu$m \cite{shiomi2006, saaskilahti2015}. 

With recent advances in computational resources, it is now becoming possible to include higher-order four-phonon processes in the prediction of phonon transport properties via the BTE-based approach \cite{feng2016, feng2018, ravichandran2020, xia2020, jain2020}. However, due to several orders of magnitude higher computational cost of four-phonon processes compared to three-phonon counterpart \cite{feng2016}, the application of four-phonon processes is currently limited to simple material systems with fewer than 8-10 atoms in the unitcell \cite{feng2016, feng2018, ravichandran2020, xia2020, jain2020, jain2022}. For the particular case of carbon-based reduced dimensionality systems, such as graphene and CNTs, this is even more challenging as the BTE is required to be solved iteratively for the correct description of Normal and Umklapp phonon processes \cite{lindsay2009}, and thus requiring evaluation of four-phonon scattering rates at each iteration step. 

In this work, the thermal transport properties of (10,0) SWCNTs consisting of 40 atoms unitcell are evaluated by considering both three- and four-phonon processes by employing the iterative solution of the BTE to address the unresolved question of the length dependence of $\kappa$ of SWCNTs.

The thermal transport in metallic/semi-metallic SWCNTs is predominantly due to atomic vibrations, i.e., phonons \cite{yamamoto2004}. The $\kappa$ of phonon is obtainable from BTE along with the Fourier law as \cite{mcgaughey2019, jain2020}: 
\begin{equation}
 \label{eqn_k}
    \kappa = \sum_i c_{ph, i} v^2 \tau_i,
\end{equation}
where the summation is over all the phonon modes in the Brillouin zone enumerated by $i\equiv(q,\nu)$, where $q$ and $\nu$ are phonon wavevector and mode index, and $c_{ph,i}$, $v$, and $\tau_i$ represent phonon specific heat, group velocity, and transport lifetime respectively. The phonon transport lifetimes are obtained from the iterative solution of the BTE by considering phonon-phonon scattering via three- and four-phonon scattering processes and phonon-boundary scattering corresponding to a characteristic length scale, $L$, as $\tau_{bdry} = {L}/{2|v|}$. Instead of phonon transport lifetimes, if phonon relaxation lifetimes are employed in Eqn.~\ref{eqn_k}, then it corresponds to the well-known relaxation time approximation (RTA) of the BTE. Further details regarding the calculation of phonon heat capacity, group velocity, and relaxation/transport lifetimes are presented in Refs.~\cite{jain2020}. The phonon modes are classified as longitudinal when participating atoms are majorly displayed along the length of the tube, i.e, $|\bold{e_i}.\hat{\bold{n}}| > 0.5$, where $\bold{e_i}$ represents eigenvector of the concerned phonon mode and $\hat{\bold{n}}$ is a unit vector along the length of the tube. The modes which are not longitudinal are classified as transverse modes.

The computation of phonon thermal conductivity via Eqn.~\ref{eqn_k} requires harmonic, cubic, and quartic interatomic force constants \cite{jain2020}. The harmonic force constants are obtained using density functional perturbation theory (DFPT) \cite{baroni2001} as implemented in planewaves-based quantum simulation package Quantum Espresso \cite{giannozzi2009} with PBEsol pseudo-potentials \cite{perdew2008} with converged electronic wavevector grid and plane wave energy cutoff of size 10 and 80 Ry respectively. The harmonic force constants are initially obtained on a phonon wavevector grid of size 10 (with 40 atoms unitcell) and are later interpolated to a finer grid of size $N_q=100$ for phonon-phonon scattering rate calculations ($80$ for four-phonon processes). The anharmonic force constants are obtained by Taylor series fitting of force-displacement data obtained from randomly displacing atoms by $0.05$ or $0.10$ $\text{\AA}$ from their equilibrium positions. Six hundred such structures, each with 200 atoms in the supercell (5 repetitions of unitcell along the tube length), are considered for the force-displacement dataset. The cubic and quartic interaction cutoffs are set at $4^{\text{th}}$ and $1^{\text{st}}$ neighbor shell. The change in $\kappa$ obtained using the relaxation time approximation (RTA) solution of BTE and using only three-phonon scattering is less than 6\% in changing the cubic interaction cutoff from $4^{\text{th}}$ to $3^{\text{rd}}$ neighbor-shell at 300 K.

\begin{figure}
\centering
\epsfbox{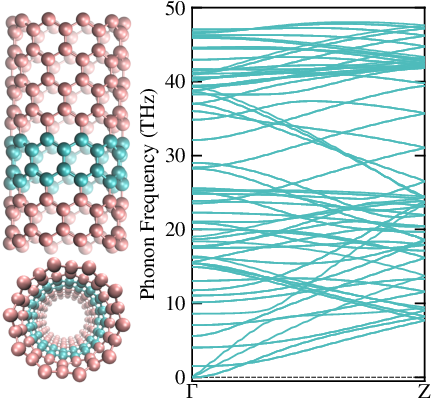}
\caption{The phonon dispersion of considered SWCNT. The structure of considered SWCNT is shown in the left and has 40 atoms in the unitcell (highlighted with cyan color).}
\label{fig_dispersion}
\end{figure}

The structure of considered SWCNT and phonon dispersion are presented in Fig.~\ref{fig_dispersion}. The considered SWCNT is an achiral zigzag (10,0) with 40 atoms in the unitcell (highlighted in cyan in Fig.~\ref{fig_dispersion}) and $7.89$ $\text{\AA}$ diameter. While the phonon dispersion of SWCNTs is discussed in detail in many studies (from theoretical and analytical/empirical forcefields) \cite{lindsay2009, bruns2020}, it is worth emphasizing that in the long wavelength limit, the flexural acoustic phonon modes (doubly degenerate) have quadratic phonon dispersion. In contrast, the twisted/rotational and longitudinal modes have linear dispersions. The phonon velocities for twisted and longitudinal modes from DFPT obtained here are $13.2$ and $21.1$ km/s compared to $13.6$ and $21.4$ reported by Bruns et al.~\cite{bruns2021} using empirical Tersoff forcefield. 

\begin{figure}
\centering
\epsfbox{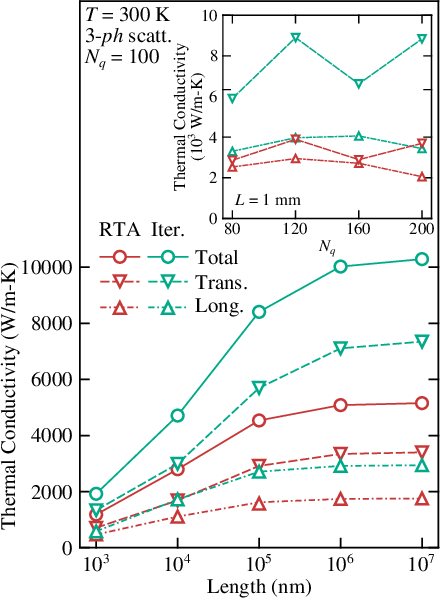}
\caption{The length-dependent thermal conductivity of SWCNT as obtained using the RTA (red) and iterative (cyan) solution of BTE by considering only three-phonon scattering. The total thermal conductivity is not converged up to 1 mm length due to the non-scattering of transverse phonon modes via Umklapp three-phonon processes. The results are obtained using a phonon wavevector grid of size $N_q=100$. The variation of longitudinal and transverse modes thermal conductivity with phonon wavevector grid is reported in the inset for a tube length of 1 mm. The contribution from transverse modes is not converged with the phonon wavevector grid for $N_q$ up to 200. }
\label{fig_k3ph}
\end{figure}

Considering only three-phonon scattering and empirical Tersoff potential, Lindsay et al.~\cite{lindsay2009} found that acoustic phonons fail to satisfy scattering conservation rules and can not undergo Umklapp scattering with other acoustic phonons. They found that using the RTA solution of the BTE, which wrongly treats population re-distributing Normal processes as resistive, the $\kappa$ of SWCNT converges with length, but when the iterative solution of the BTE is employed to correctly describe Normal and Umklapp processes, $\kappa$ increases more than two-fold and fails to converge with the tube length. Similar findings are recently reported by Bruns et al.~\cite{bruns2020}, where authors reported divergence of transverse acoustic phonon lifetimes (flexural and twisted/rotational acoustic modes) in the long wavelength limit. In this work, using only three-phonon scattering, the $\kappa$ obtained using DFT force constants converges to 5200 W/m-K using the RTA solution of the BTE (Fig.~\ref{fig_k3ph}) and it compares well with the corresponding value of 6200 W/m-K reported by Lindsay et al.~\cite{lindsay2009} using the Tersoff forcefield. Further,  the $\kappa$ obtained here using the iterative solution of the BTE is more than a factor of two higher than that obtained via the RTA solution and stays non-converged even for tube lengths of 1 mm ($\kappa$ for 1 mm tube is 19\% higher than that for 100 $\mu$m tube length). The mode-dependence of total-$\kappa$ in Fig.~\ref{fig_k3ph} shows that while the longitudinal modes contribution converges well with tube length, the transverse modes contribution is non-converged and keeps increasing with tube length. The inset of Fig.~\ref{fig_k3ph} further highlights this non-scattering of transverse modes via Umklapp three-phonon processes, where $\kappa$ contributions from both longitudinal and transverse modes are converged with phonon wavevector grid at the RTA level solution of the BTE (treating Normal processes also as resistive processes) but the contribution of transverse modes is not converged with phonon wavevector grid with the iterative solution of the BTE even for grid sizes of 200. 

\begin{figure}
\centering
\epsfbox{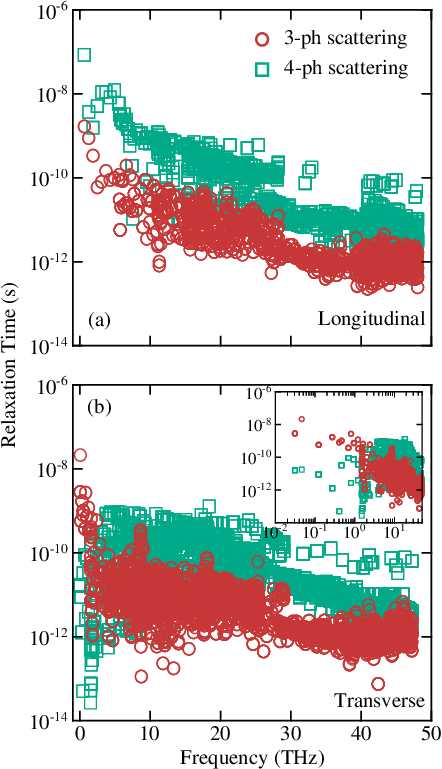}
\caption{The phonon relaxation times obtained by considering three-phonon (red) and four-phonon (cyan) scattering for (a) longitudinal and (b) transverse phonon modes. The low-frequency transverse phonon modes are unscattered via three-phonon processes but undergoes scattering via four-phonon processes. The inset in (b) is same as (b) on the log-frequency scale.  }
\label{fig_tau}
\end{figure}
The effect of four-phonon scattering on the phonon relaxation lifetimes (not transport lifetimes) is reported in Figs.~\ref{fig_tau}(a) and \ref{fig_tau}(b) for longitudinal and transverse phonon modes. For longitudinal modes, the phonon scattering via four-phonon processes is weaker than that via three-phonon processes for all phonons. The same applies to transverse modes with frequencies larger than 2-3 THz. In contrast, when considering low-frequency transverse modes, the situation is reversed, with four-phonon processes significantly surpassing three-phonon processes in terms of phonon scattering strength by several orders of magnitude, i.e., the long wavelength phonon modes which are otherwise unscattered via three-phonon processes undergoes scattering with four-phonon processes.

\begin{figure}
\centering
\epsfbox{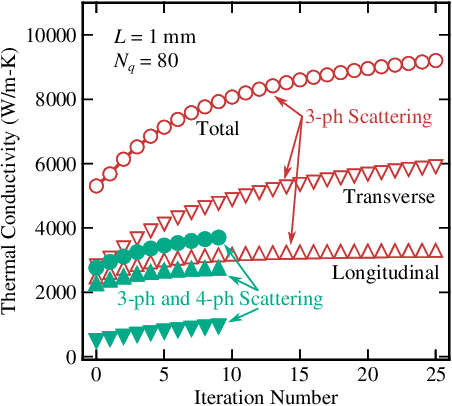}
\caption{The effect of four-phonon scattering on the thermal conductivity of SWCNTs. The thermal conductivities are obtained using the iterative solution of the BTE with iteration 0 corresponding to the RTA solution of BTE. With the inclusion of four-phonon scattering, while longitudinal modes remain unaffected, the contribution of transverse modes decreases drastically. The contribution of transverse modes to the total thermal conductivity decreases from 64\% to 25\% with the inclusion of four-phonon scattering.}
\label{fig_4ph_iteration}
\end{figure}

The effect of four-phonon scattering on the predicted $\kappa$ of SWCNTs using the iterative solution of the BTE is reported in Fig.~\ref{fig_4ph_iteration}. Iteration 0 in Fig.~\ref{fig_4ph_iteration} corresponds to the RTA solution of the BTE. As can be seen from the figure,  the longitudinal modes contribution remains largely unaffected with the inclusion of four-phonon scattering. The contribution of transverse modes, however, reduces drastically: using the RTA solution of the BTE, the contribution of transverse mode reduces from more than 2800 W/m-K to less than 500 W/m-K with four-phonon scattering. This reduction in $\kappa$ of transverse modes originates from Umklapp four-phonon processes, as even with the iterative solution of the BTE, the contribution of transverse modes remains lower than 1000 W/m-K (compared to around 6000 W/m-K with only three-phonon scattering). 

It is worth mentioning that the phonon wavevector grid employed here for four-phonon scattering rates calculation is of size 80, which results in 9600 phonon modes. This count is the same as that reported recently for graphene by Han and Ruan ~\cite{han2023}, where authors found convergence of three- and four-phonon scattering limited $\kappa$ for 9600 phonon modes but non-convergence of three-phonon limited $\kappa$ even with 345600 phonon modes. Further, this study does not include the effect of temperature renormalization on interatomic force constants arising from atomic thermal displacements. As is shown recently for graphene \cite{alam2023}, the thermal stochastic technique \cite{hellman2013}, which is the method of choice to capture this effect, is not applicable for materials with quadratic phonon dispersions. 


In summary, the thermal conductivity of single wall carbon nanotube is evaluated numerically using the iterative solution of the Boltzmann transport equation by considering both three- and four-phonon scatterings with density functional theory-based harmonic, cubic, and quartic interatomic force constants. The flexural and twisted phonon modes, which remain unscattered via Umklapp three-phonon processes, are found to undergo Umklapp scattering with four-phonon processes, resulting in a non-ballistic transport with predicted thermal conductivity of 3700 W/m-K at 300 K.

{\bf Acknowledgement}
The author acknowledge the financial support from National Supercomputing Mission, Government of India (Grant Number: DST/NSM/R\&D-HPC-Applications/2021/10) and Core Research Grant, Science \& Engineering Research Board, India (Grant Number: CRG/2021/000010).  The calculations are carried out on SpaceTime-II supercomputing facility of IIT Bombay and PARAM Sanganak supercomputing facility of IIT Kanpur.

{\bf Data Availability}
The raw/processed data required to reproduce these findings is available on a reasonable request via email.


%

\end{document}